\documentstyle[twoside,epsfig] {article}

\catcode`\@=11
\long\def\@makefntext#1{
\protect\noindent \hbox to 3.2pt {\hskip-.9pt  
$^{{\eightrm\@thefnmark}}$\hfil}#1\hfill}		

\def\@makefnmark{\hbox to 0pt{$^{\@thefnmark}$\hss}}	
	
\def\ps@myheadings{\let\@mkboth\@gobbletwo
\def\@oddhead{\hbox{}
\rightmark\hfil\eightrm\thepage}   
\def\@oddfoot{}\def\@evenhead{\eightrm\thepage\hfil
\leftmark\hbox{}}\def\@evenfoot{}
\def\sectionmark##1{}\def\subsectionmark##1{}}

\oddsidemargin=\evensidemargin
\newcounter{sectionc}\newcounter{subsectionc}\newcounter{subsubsectionc}
\renewcommand{\section}[1] {\vspace{12pt}\addtocounter{sectionc}{1} 
\setcounter{subsectionc}{0}\setcounter{subsubsectionc}{0}\noindent 
	{\tenbf\thesectionc. #1}\par\vspace{5pt}}
\renewcommand{\subsection}[1] {\vspace{12pt}\addtocounter{subsectionc}{1} 
	\setcounter{subsubsectionc}{0}\noindent 
	{\bf\thesectionc.\thesubsectionc. {\kern1pt \bfit #1}}\par\vspace{5pt}}
\renewcommand{\subsubsection}[1] {\vspace{12pt}\addtocounter{subsubsectionc}{1}
	\noindent{\tenrm\thesectionc.\thesubsectionc.\thesubsubsectionc.
	{\kern1pt \tenit #1}}\par\vspace{5pt}}
\newcommand{\nonumsection}[1] {\vspace{12pt}\noindent{\tenbf #1}
	\par\vspace{5pt}}

\newcommand{\textlineskip}{\baselineskip=13pt}
\newcommand{\smalllineskip}{\baselineskip=10pt}

\def\eightcirc{
\begin{picture}(0,0)
\put(4.4,1.8){\circle{6.5}}
\end{picture}}
\def\eightcopyright{\eightcirc\kern2.7pt\hbox{\eightrm c}} 

\newcommand{\copyrightheading}[1]
	{\vspace*{-2.5cm}\smalllineskip{\flushleft
        {\footnotesize ICRA Letters, 29 July 2000\\
        $\eightcopyright$\, World Scientific Publishing Company
        }\\
	 }}

\def\abstracts#1#2#3{{
	\centering{\begin{minipage}{4.5in}\baselineskip=10pt\footnotesize
	\parindent=0pt #1\par 
	\parindent=15pt #2\par
	\parindent=15pt #3
	\end{minipage}}\par}} 

\newcommand{\bibit}{\nineit}

\renewenvironment{thebibliography}[1]
	{\frenchspacing
	 \ninerm\baselineskip=11pt
	 \begin{list}{\arabic{enumi}.}
        {\usecounter{enumi}\setlength{\parsep}{0pt}     
	 \setlength{\leftmargin 12.7pt}{\rightmargin 0pt} 
         \setlength{\itemsep}{0pt} \settowidth
	{\labelwidth}{#1.}\sloppy}}{\end{list}}

\newcounter{itemlistc}
\newcounter{romanlistc}
\newcounter{alphlistc}
\newcounter{arabiclistc}

\def\@citex[#1]#2{\if@filesw\immediate\write\@auxout
	{\string\citation{#2}}\fi
\def\@citea{}\@cite{\@for\@citeb:=#2\do
	{\@citea\def\@citea{,}\@ifundefined
	{b@\@citeb}{{\bf ?}\@warning
	{Citation `\@citeb' on page \thepage \space undefined}}
	{\csname b@\@citeb\endcsname}}}{#1}}

\newif\if@cghi
\def\cite{\@cghitrue\@ifnextchar [{\@tempswatrue
	\@citex}{\@tempswafalse\@citex[]}}
\def\citelow{\@cghifalse\@ifnextchar [{\@tempswatrue
	\@citex}{\@tempswafalse\@citex[]}}
\def\@cite#1#2{{$\null^{#1}$\if@tempswa\typeout
	{IJCGA warning: optional citation argument 
	ignored: `#2'} \fi}}

\def\@refcitex[#1]#2{\if@filesw\immediate\write\@auxout
	{\string\citation{#2}}\fi
\def\@citea{}\@refcite{\@for\@citeb:=#2\do
	{\@citea\def\@citea{, }\@ifundefined
	{b@\@citeb}{{\bf ?}\@warning
	{Citation `\@citeb' on page \thepage \space undefined}}
	\hbox{\csname b@\@citeb\endcsname}}}{#1}}

\def\@refcite#1#2{{#1\if@tempswa\typeout
        {IJCGA warning: optional citation argument
	ignored: `#2'} \fi}}

\def\refcite{\@ifnextchar[{\@tempswatrue
	\@refcitex}{\@tempswafalse\@refcitex[]}}

\def\pmb#1{\setbox0=\hbox{#1}
	\kern-.025em\copy0\kern-\wd0
	\kern.05em\copy0\kern-\wd0
	\kern-.025em\raise.0433em\box0}

\def\fnt#1#2{\footnotetext{\kern-.3em
	{$^{\mbox{\scriptsize #1}}$}{#2}}}

\font\tenrm=cmr10
\font\tenit=cmti10 
\font\tenbf=cmbx10
\font\bfit=cmbxti10 at 10pt
\font\ninerm=cmr9
\font\nineit=cmti9

\font\eightrm=cmr8

\def\qed{\hbox{${\vcenter{\vbox{			
   \hrule height 0.4pt\hbox{\vrule width 0.4pt height 6pt
   \kern5pt\vrule width 0.4pt}\hrule height 0.4pt}}}$}}

\begin{document}



\normalsize\textlineskip
\thispagestyle{empty}
\setcounter{page}{1}


\vspace*{0.88truein}

\centerline{\bf DARBOUX COSMOLOGICAL FLUIDS IN COMOVING TIME}
\vspace*{0.035truein}
\vspace*{0.37truein}
\centerline{\footnotesize RENATO KLIPPERT$^{1,3}$, HARET C. ROSU$^{2,3}$, REMO RUFFINI$^{3}$}
\vspace*{0.015truein}
\centerline{\footnotesize\it $^1$ Brazilian Center for Research in Physics,
R. Dr.\ Xavier Sigaud 150 Urca, 22290-180 Rio de Janeiro RJ, Brazil}
\baselineskip=10pt
\centerline{\footnotesize\it $^{2}$ Instituto de F\'{\i}sica,
Universidad de Guanajuato, Apdo Postal E-143, Le\'on, Gto, Mexico}
\centerline{\footnotesize\it $^{3}$ University of Rome ``La Sapienza'' and ICRA, I-00185 Rome, Italy}

\vspace*{10pt}
\vspace*{0.225truein}

\vspace*{0.21truein}
\abstracts{The barotropic indices and the corresponding FRW scale factors of the so-called Darboux cosmological fluids are presented in the comoving time axis, which is 
the natural one for the phenomenology related to the cosmological data. Some useful comments
on the features of the plots are included.
}{}{}


\textlineskip                  
\vspace*{12pt}                 

\vspace*{1pt}\textlineskip	
\vspace*{-0.5pt}
\noindent


\noindent




\noindent

In a previous work, an interesting class of Darboux cosmological fluids in closed and
open FRW
models of the $\Lambda =0$ cosmology has been introduced by one of the authors.$^{1}$ However, the results have been
displayed in the conformal time axis, whereas for comparison with definite cosmological data the 
comoving time axis is usually needed. The purpose of this letter is to present the 
cosmological comoving evolution of that class of fluids and add several heuristic comments 
on the obtained results. We plot here the comoving time dependent barotropic indices and the 
corresponding scale factors for open FRW universes,$^2$ respectively.  
Our general conclusions are presented in the following.

(i) For the open cases, we find that $\gamma _{-1}(t,\lambda)$ displays a single deep structure where an 
accelerating universe can exist, $\gamma_{-1}(t,\lambda)<2/3$. Moreover, depending on the value of the $\lambda$ parameter,$^1$ a 
negative region of limited extent may occur indicating the presence of a Chaplygin-like fluid whose origin may be 
due to the contribution of quantum field extended objects, such as d-branes.$^{3}$ 

(ii) For the closed, matter- and radiation-dominated cases, we find a more complicated damped periodic $\gamma _{+1}(t,\lambda)$, for which the accelerating region is present only in the first period.
No Chaplygin-like region occurs in these cases, which points to more stable cosmological conditions. The vacuum-dominated closed case has a Chaplygin-like (unstable) oscillatory behavior forever.

A very definite prediction of the 
model is that there is a single localized accelerating region (SLAR) for all types of non flat FRW universes, except for the closed vacuum case. Concerning a possible application to cosmological data,$^{4}$ this means
that only the sample of supernovae with redshifts corresponding to the SLAR can show a clear-cut accelerating effect. 
Such behavior can be tested by means of a precise calibration of the supernovae cosmological data, which should take into account the determination of the time scale $T$ introduced in the plots.







\begin{figure}[htbp]
\leavevmode
\centerline{
\centering
\epsfxsize=80ex
\epsfbox{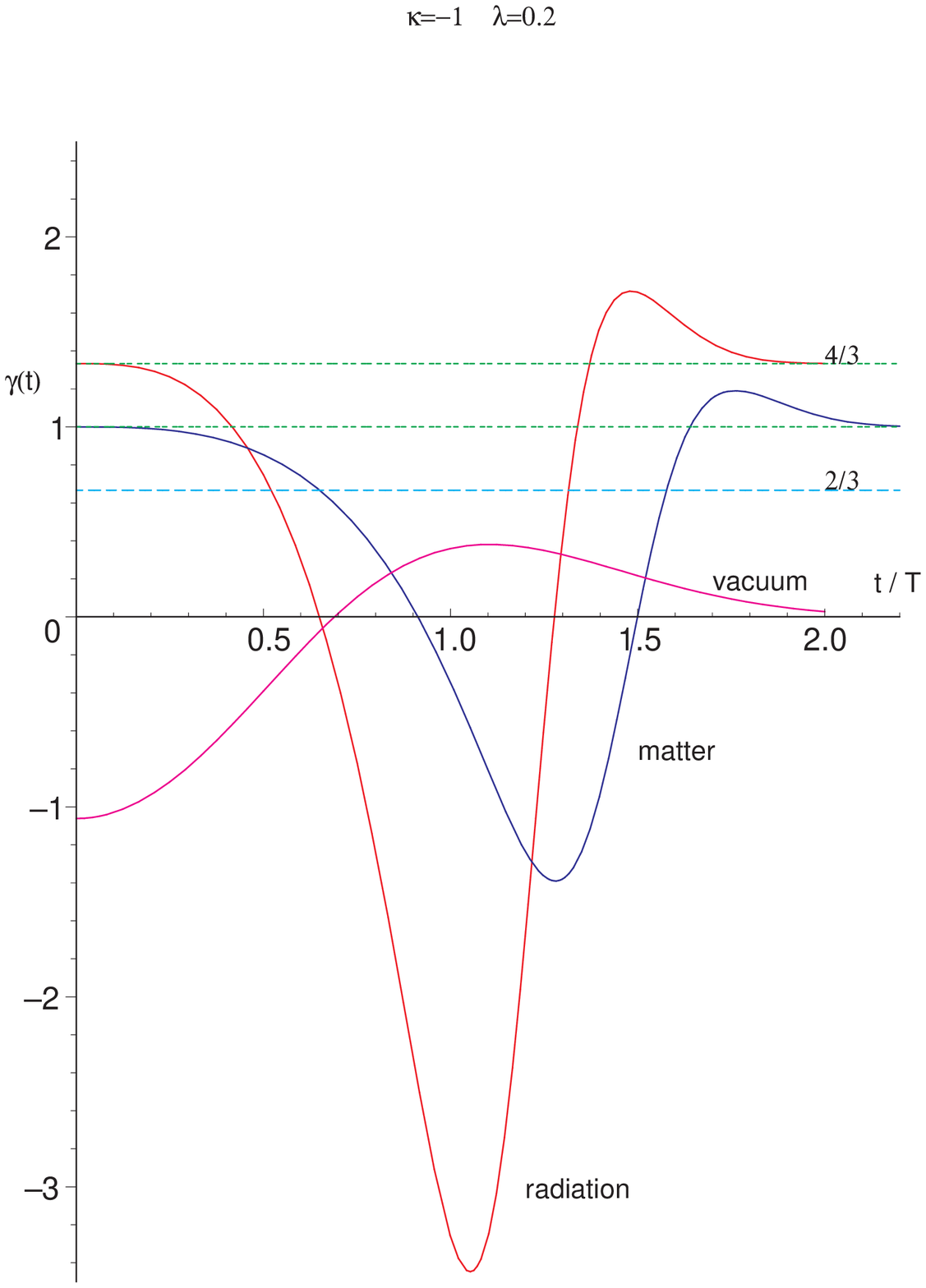}}
\caption{The time-dependent $\gamma _{-1}$ index for three classes of fluids we are interested in: vacuum, radiation and matter.  The plot corresponds to an open universe of $\lambda =0.2$ along the comoving time axis. There is a small region associated to each $\gamma _{-1}$ where it is negative (Chaplygin-like fluid).  }
\protect\vspace{1\baselineskip}
\label{fig1}
\end{figure}

\begin{figure}[htbp]
\leavevmode
\centerline{
\centering
\epsfxsize=80ex
\epsfbox{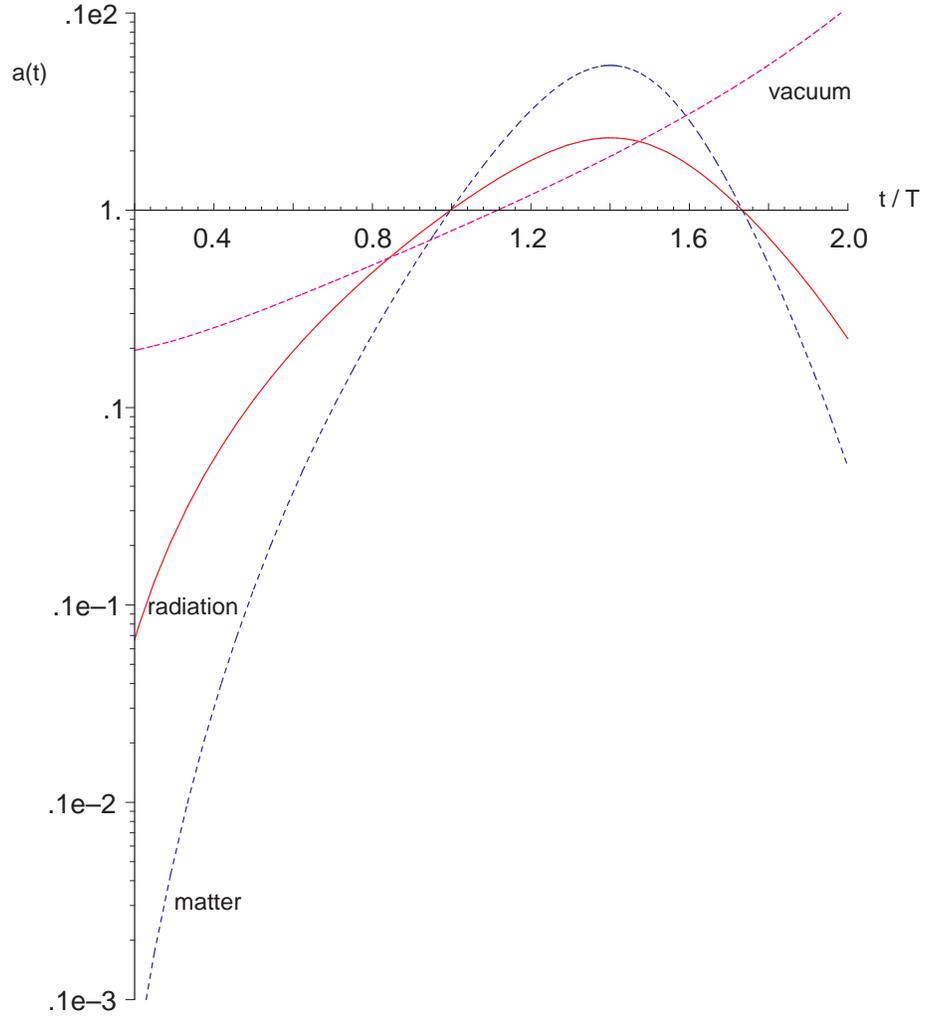}}
\caption{The scale factors corresponding to the previous cases.  The universe undergoes a vacuum-dominated region followed by a radiation epoch, then by a matter era and finally ends in a vacuum phase.}
\protect\vspace{1\baselineskip}
\label{fig2}
\end{figure}




\clearpage
\nonumsection{Acknowledgements}
\noindent
RK would like to acknowledge Brazilian CAPES foundation for a grant. 


\nonumsection{References}


\end{document}